# Digital audiovisual archives in the humanities: problems and challenges

Peter Stockinger
Maison des Sciences de l'Homme
Paris - Chania 2003

## Introduction

This paper presents:
– the audiovisual archive program in humanities (social and human sciences) launched in 2001 by the Maison des Sciences de l'Homme (MSH) in Paris;
– a working (indexing and annotation) environment for individual or collective users ("communities") to extract audiovisual segments, to classify them in personal archives, to describe them following the particular user point of views and to reuse them in given academic and educational, viz. professional activities.

The goal of the presentation and discussion of the audiovisual archive program in humanities is to understand better the interests and challenges of audiovisual archives in general for *public* research and educational institutions.

## 1) The audiovisual archives of the MSH

The Maison des Sciences de l'Homme (MSH) is a public research foundation, located in Paris (France)[1].

Since 2001, it has started to build an audiovisual archive in social and human sciences. The principal – institutional – goals of this programare the following ones:

*1) Production of scientific knowledge in the humanities*
– constitute an (audiovisual, textual, etc.) content base that could play the role of an "incubator" for the emergence of novel knowledge as well as for collectively shared knowledge standards,
– create an observatory for ongoing research and research trends.

*2) Conservation of scientific knowledge*
– preserve and conserve the scientific contributions of researchers, especially in the humanities,
– create a public and open archive of scientific heritage in humanities.

*3) Diffusion of specialized and very specialized knowledge in the humanities*
– create central accessto sometimes highly dispersed information and knowledge produced by a multiplicity of small and small research structures (labs, centers, associations, scientific programs, etc.) in the humanities,
– create a base of reusable elements for multi-support and "personalized" publishing of scientific knowledge in the humanities on the web, by the means of DVD, CD-ROM, VHS cassettes, etc.

*d) task-oriented exploitation of scientific knowledge in the humanities*
– create a sharable knowledge base for research and teaching communities,
– create high-grade knowledge references for professional uses (media, policymakers, etc.),
– create a "competence pool" for young researchers.

The principal content of the audiovisual archives of the MSH is footage of:
1. research seminars;
2. symposia, workshops;
3. the "work and life" in research labs;
4. specific and often highly specialized competencies and skills as well as concrete realizations (publications, expositions, artifacts such as tools and instruments, etc.);
5. and *especially* in-depth interviews with researchers (average time of one interview session: 2, 5 hours; a researcher may repeat as often as he wants an interview session; average time of one interview: 3.5 hours)

The actual state of the art of the archive can be summarized as follows:
– *simple access* to these archives via the web site of the MSH (http://www.msh-paris.fr – section "Manifestations En Ligne"; cf. figure 1)[2];

---

[1] Website of the Maison des Sciences de l'Homme : http://www.msh-paris.fr

[2] This access ceased to function in 2016. To access and explore the AAR scientific heritage, please use this link: http://www.archivesaudiovisuelles.fr/fr/

- *rich access*, via the portal of the archive – the "Médiathèque Sciences Humaines et Sociales"- (http://e-msha.msh-paris.fr/Mediatheque_SHS; cf. figure 2)[3];
- 450 hours of scientific video on-line (some 300 hours have yet to be digitized until the end of August 2003);
- most of the videos are in three formats: 56k; 256k and MPEG 1 or 2;
- the major part of the content is in French (but there does not exist any limitation for the use of other languages …).

The principal *information unit* of the audiovisual archive is the so-called "**event**" (scientific, educational, etc.):it
- is an interview (one session of an interview), a seminar, a workshop, a round table, a presentation, or demonstration, …;
- is generated as a dynamic website composed by a set of data registered in a relational database (MS SQL server);
- Is the *base indexing and description unit* in the media portal (implemented on MS Sharepoint server; cf. Figure 3).

The *dynamic website* visualizing a scientific event is organized in :
- a "home page" (cf. figure 4) describing the event and the actors of it and providing access to the other pages;
- a "video page," which is the principal interface for consulting an event (cf. figure 5);
- a "thematic index page" with access to the principal scientific notions and their definitions as well as to the videos where they are developed (cf. figure 6);
- several other pages containing supplementary information on the event (related documents, …).

The "video page" (cf. figure 5) contains :
1. the embedded MS media player;
2. the principal thematic sections following which an event can be explored;
3. supplementary information sources (as pdf, ppt, image, other video… files) that accompany an event;
4. some information concerning a selected video file (title, duration, format, …).

---

[3] This access ceased to function in 2016. To access and explore the AAR scientific heritage, please use this link: http://www.archivesaudiovisuelles.fr/fr/

Finally, the "canonical" production and editing chain adopted for realizing an event can be summarized as follows:
1. shooting of an event (with DV cameras);
2. digitizing of footage;
3. cutting of a digitized footage in several avi files;
4. post-editing of each file;
5. thematic description and indexing of the files;
6. exporting of *avi* files in *asf* and *mpeg* formats;
7. registration of *asf* and *mpeg* files as well other files and information in the database;
8. generation of a dynamic website of the event;
9. broadcasting of the website within the MSH archive.

## 2) Users, objectives,and needs

The principal users of the MSH audiovisual archive in humanities are:
- researchers and research teams;
- teachers and graduate students;
- professional user groups (journalists, policymakers, etc.).

As already explained above (cf. 1st chapter) the archive is supposed to fulfil:
- a heritage function
- an observatory function,
- a transfer (teaching and training) function,
- an incubating function for research and knowledge production.

To satisfy genuinely these functions, the principal needs may be identified as follows:. Under the "umbrella" of the MSH audiovisual archive in social and human sciences, it should be possible:
- to create **work spaces** for individual and collective users ("communities") attuned to the specific (scientific, educational, professional) usages of a set of – scientific - events;
- (for authorized users) to identify and to select, within the existing audiovisual data pool, **(audiovisual and textual) segments** that are relevant regarding the users' knowledge and information requirements;
- (for individual or collective users) to specify their **viewpoints** following which the selected segments should be classified and described;
- (for individual or collective users) to create his/her **"personalized" ontologies (thematic grids, thesaurus, etc.)** and **annotation categories** to index and annotate (describe) the selected segments following a chosen and specified viewpoint;

- to **share** and (collectively) **update** viewpoints, ontologies, and annotation categories;
- to create **personal archives** (or again *archives* attuned to group specific interests) of scientific events out of the common MSH research archive;
- to fully **research** and **explore** video and textual segments indexed and described following a specific view point;
- (for an individual or collective user) to produce a **new "montage"** (a new "hyper-document") based on the selected and described segments
- **to edit** such a **montage** either as a website, a CD-ROM, a DVD, a VCD, etc.

## 3) The OPALES Project

These needs constitute, globally speaking, the "input" and certainly, one of the principal motivation of the French OPALES project.

Opales (« **O**utils pour des **P**ortails **A**udiovisue**l**s **E**ducatif et **S**cientifique ») has been a French R&DT project, financed by the Ministery of Industry within a specific national program (the RIAM - **R**echerche et **I**nnovation en **A**udiovisuel et **M**ultimedia - program) aiming at the development of innovative technological, economic and socio-cultural models in the digital audiovisual sectors.

The Opales project has been leaded by the INA (Institut National de l'Audiovisuel). The principal partners have been:
- Université de Montpellier – LIRMM (Laboratoire d'Informatique, de Robotique et de Microélectronique de Montpellier) ;
- Université d'Angers ;
- Ministère de l'Education National – CNDP (Centre National de Documentation Pédagogique) ;
- Cité des Sciences de la Villette à Paris ;
- France 5 ;
- Maison des Sciences de l'Homme (MSH) – Equipe Sémiotique Cognitive et Nouveaux Medias (ESCoM) ;
- Etc.

The « output » of the project is the prototype of an integrated environment (the **Opales – environment**) for describing (indexing, annotating, …) selected audiovisual and textual segments following a specific viewpoint as well as for producing a « new montage » of described segments.

The project started in September 1999 and finished in March 2003. A second phase of this project is planned to overcome the limitations of the actually existing prototype (cf. chapter 6).

## 4) The Opales indexing and annotation environment

The general architecture of the Opales environment (cf.figure 7; figure 8) is composed of:

1. the *Opales client*;
2. the *video server* and
3. the *Opales environment server* (cf.figure 9).

The Opales client interface is composed of two components (cf.figure 8):

1. A research and exploration interface;
2. A working interface.

The **working interface** is composed by a set of specialized tools such as:
1. a **video explorer tool** (an enhanced MS media player for selecting segments and zooming on selected zones in an audiovisual stream);
2. an **ontology builder** (of concepts or themes and relations between concepts and themes representing the knowledge of a given domain of expertise);
3. a **conceptual graph editor** (which takes as the input an ontology of themes and relations and which produces as the output a particular *conceptual or thematic configuration*, i.e., a set of selected themes and selected relations between themes representing specific scenes in an audiovisual segment);
4. a **point of view editor** (which is a very generic "formulary" composed of a set of features and which can be attuned to the requirements of a user or user group);
5. a **video segment indexing and annotation interface** (by the means of which themes, conceptual graphs, viewpoints, or simply free annotations are "attached" to a video segment or a particular zone within a segment);
6. a **segment montage editor** (allowing mainly to create oriented navigation paths through a set of selected and described video segments);
7. a **tool** for creating a personal or collective **work space**.

Note: the working interface is an open one – other services can be added to the already existing ones.

The **research and exploration interface** offers the user different possibilities to access the content of a (personal or collective) archive:
- direct access to the personal library of selected video segments;
- access to specific scenes developed in selected segments via a request by the means of conceptual graphs;
- access to the content of a personal library of video segments via defined themes in an ontology;
- access via keywords to the personal library of selected segments;
- access to the hyper-documents composed out of selected and described (indexed and annotated) video segments.

## 5) A typical user scenario

*A (not fictive) case:* A researcher in social history more especially concerned by the processes of western industrialization wants to build his/her personal library of audiovisual and textual documents for different purposes:
- evaluation and critical comparison of existing knowledge regarding further research;
- identification of specific research topics;
- preparation of a series of specialized workshops, research seminars, or publications;
- ….

To assist the researcher in these and other comparable activities, the Opales environment has been conceived and developed (as a prototype) about **8 very typical steps** (or "tasks") that cover – as it seems – most of the above quoted researcher's activities. Naturally, in speaking of "steps" or "tasks," the Opales environment refers to an ideal scenario : for his/her concrete needs and goals, a researcher may only use very partially the possibilities offered by the Opales environment.

**Step 1)** Once connected to the Opales interface, the researcher opens the **Opales work space tool** and creates his own (personal) work space in which he will have :
- the Opales video description environment;
- and more or less publicly resources such as already existing ontologies, viewpoints, commented video segments, etc.

**Step 2)** The researcher explores – via the **MSH audiovisual archive portal**[4] - the for his research topic relevant scientific events (seminars, interviews,…) and creates (in the form of "annotated bookmarks"), progressively, his personal base library of relevant scientific events (this base can be publicly accessible or accessible only for a particular user group).

**Step 3)** Then, the researcher starts by identifying, within the audiovisual data composing an event, the *segments* that are of particular interest for him and creates his personal library ("list") of (audiovisual, textual, …) segments (this activity can, naturally, be undertaken repeatedly, …). The activity of selecting specific segments, the researcher can perform with the help of the **enhanced MS Media Player** that possesses several options, such as the identification of start and end points of a segment, zooming on specific zones of a single frame, etc.

**Step 4)** The researcher starts by specifying his "ontology." This – crucial – step can be, normally, decomposed in several tasks:
1. specification of a hierarchy of themes (or concepts) representing the knowledge of a given domain (i.e., in our case, modern industrialization processes);
2. specification of typical relationships between themes (apart from the *is-a* relationship, which is the default relationship for building an ontology of themes and relationships);
3. elaboration of definitions for each theme and each relationship previously identified;
4. building the ontology with the help of the **Opales ontology builder**;
5. assessment of the ontology.

It has to be stressed that the specification of a hierarchy of themes is a very complicated process. It can be observed that this process rather typically proceeds following the epistemological distinction between several main types of themes, such as:
a) *notional themes* (i.e., themes referring to a given research field: actors, objects, beliefs and theories, etc.);
b) *rhetorical themes* (i.e., themes referring to the discourse activities employed for "speaking" (arguing, describing, narrating, refuting, etc.)

---

[4] cf. http://e-msha.msh-paris.fr/Mediatheque_SHS. This access ceased to function in 2016. To access and explore the AAR scientific heritage, please use this link: http://www.archivesaudiovisuelles.fr/fr/

about an object (activity, etc.) in a research field;
c) *contextual themes* (i.e., themes referring to the context, the history, and tradition, the objectives, the resources, the practical exploitations and challenges, etc. of a research activity).

If more complicated is the process of specifying an ontology of thematic relations that should define typical configurations between a set of selected themes (cf. above the work with the conceptual graph editor). Indeed, recurrently used are the following types of thematic relations:
a) *classifications* (such as the taxonomic and the mereonymic "part-whole" relationship);
b) *practical inferences* (such as causal relationships, intentional ones, contractual ones, etc.)
c) *epistemic inferences* (such as relationships depicting "topical" or analogical reasoning patterns, etc.);
d) *modalisation* and *grading* (such as relationships for depicting preferential patterns, etc.).
e) *localizations* (such as temporal and spatial relationships).

Given the conceptual complexity which seems to be obligatorily implied in the elaboration of any relevant ontology, it is easily understandable that researchers are reluctant to get involved in this task.

In any case, a researcher at least will produce at least a simple version of an ontology of his research domain – simple version which will be more or less similar to a thematic index or again a (small) thesaurus).

Once the ontology established, the researcher can start to index his segments (parts of his segments). For this:
– he opens a segment with the enhanced MS Media Player;
– identifies the temporal positions of the parts within the segment;
– "attaches" one or more themes to these parts.

**Step 5)** After having built his ontology, the researcher, normally will have to specify **typical scenes** that are developed in the previously selected audiovisual segments.

As far as the interviews with researchers are concerned, such typical scenes are, for instance:

– the discussion of given problem (object, historical data, …) by the interviewed researcher;
– the presentation of the historical filiation (genealogy) of a given (problem, object, …);
– the qualification of already existing research;
– the qualification of the achievement of a personal research;
– the argumentation of a scientific affirmation (hypothesis, theory, …);
– the confrontation (critical comparison) of two or more research results;
– etc.

To capture and describe (index) such typical scenes in his personal library of selected audiovisual segments, the researcher:
– opens the conceptual graph editor;
– imports themes and relations from his ontology in the conceptual graph editor;
– builds conceptual graphs (i.e., configurations) of a given set of themes and relations in qualifying the particular position between two themes within a given configuration by the means of a specific, oriented, and named relation.

Each conceptual graph represents, by definition, a *type of scene* in the audiovisual segments.

After having finished the building of conceptual graphs, the researcher opens (with the enhanced MS media player) an audiovisual segment and describes the scenes developed in this segment with the help of a conceptual graph:
– he identifies the temporal positions of a given scene;
– selects a conceptual graph that represent the type of this scene;
– "attaches" the conceptual graph to this scene;
– adds keywords or textual descriptions to the instantiated graph (i.e., attached to a given scene) in using, for this, the referent field of a conceptual graph (remember, a graph has a *conceptual field* and a *referent field*, i.e., a field to which a concept refers: [CONCEPT: REFERENT]).

Note: Another – complementary – exploitation of a conceptual graph built based on an already existing ontology is to improve the search mechanism of the ontology-based search and exploration interface of Opales.

**Step 6)** To adapt ("personalize") even more the indexing and description interface, the researcher will open the **View point editor.** This editor is a highly generic formula provided by a set of already given features that the researcher restricts to his special case. He also may define new features, not yet provided by the formulary.

For instance, the researcher would like to include in his description of audiovisual segments specific moments highly important for his research such as:
– the rhetorical nature of an interview
– the relative importance of a theme developed within a segment or one of its parts;
– the credibility of information communicated within a segment;
– the "added value" of communicated information;
– the specialization degree of the communicated information;
– etc.

All these items, the researcher can specify in the view point editor and elaborate a highly specialized, "personal" description model that allows him to enhance a description of a video segment done by the sole means of an ontology and a set of conceptual graphs.

**Step 7)** The researcher may want to produce a "complex" hyper-document out of all selected and described video segments (as a .ppt presentation, for instance). For this, he uses the **video montage editor**.

**Step 8)** Finally the researcher certainly wants to decide which of his resources (ontologies, conceptual graphs and view points) could be used by other researchers or not. He also will decide which descriptions and annotations are completely personal (only for him), which others are accessible for identified user groups and which are publicly accessible. This, the user can perform with the already mentioned (cf. step 1) **personal work space tool**.

Note: There are again some other tools included in the Opales working environment, but the above-mentioned can be considered to be the most crucial ones for a scientific (pedagogical or professional) work on the data of audiovisual archives.

## 6) Difficulties, limitations, and further orientations

The Opales project has been a crucial one for the development of tools and services allowing specific user groups to exploit but also to enhance and enrich the audiovisual archives of the MSH in the humanities.

It also has been interesting to better identify difficulties in the use of such rather complex and sophisticated tools and services.

Indeed, an important number of these difficulties concern, especially the **working habitudes** and **traditions** of researchers (teachers, professional users, etc.). Besides the – well known – fact that researchers are not always disposed to share not-definitive results of their ongoing research, many of them are indeed not well-prepared to use *Opales-like* tools and services because of methodological reasons.

For instance, even if a researcher always uses some "practical" classification and description schemas, he/she is not necessarily prepared to make explicit these practical and intuitively used schemas. Working with an ontology or one or several view points, nevertheless presupposes this elicitation process of practical classification and description schemas, i.e., of schemas grounded on the experience of the researcher and tacitly used by him or his community.

In other words: such schemas constitute a part of the *scientific culture* of a research community and their elicitation is a work that is rather similar to the analysis and the interpretation of texts or other documents for which particular methodological knowledge and skills (such as semiotics or rhetoric) are needed - methodological knowledge and skills which a given research community not always may possess.

This first types of difficulties also explains another types of difficulties because the Opales working interface is provided with somehow too rich and to "general" tools and services: the "common" researcher risks to get lost in front of such an environment (and one can guess that this should also be the case for teachers and other categories of professional users).

This means that in a second phase of the Opales project, an important effort will have to be done:
– to provide simplified and "standard" working interfaces attuned to more specific but already prior identified types or profiles of working and exploitation contexts (such as "realize a course, "build a personal archive, "create a web portal, etc.);
– to provide the ontology editor, the conceptual graph editor and the view point editor with *already pre-defined models* (pre-defined

ontologies, pre-defined typical scenes developed in a video of an interview or a seminar, pre-defined view points for analyzing a segment of a scientific event, etc.).

Other difficulties concern more particularly the actual limitations of the Opales working environment. The plan is to remove these limitations in the second phase of the Opales project (and in other, related projects):
— transformation of the Opales tools into web-based tools, viz., in genuine "web services";
— introduction of a genuine multilingual dimension in these tools;
— enhancement of specific tools of the Opales environment (such as, for instance, the "montage editor for producing hyper-documents out of selected and described audiovisual segments);
— development of genuine multi-support publishing tool (for CD-ROM, DVD, websites but also traditional VHS cassettes);
— and, especially, integration of principal standards in an Opales description and annotation of audiovisual segments (standards such as DC, OAI, MPEG 7, TVA, etc.).

# Annex : Illustrations

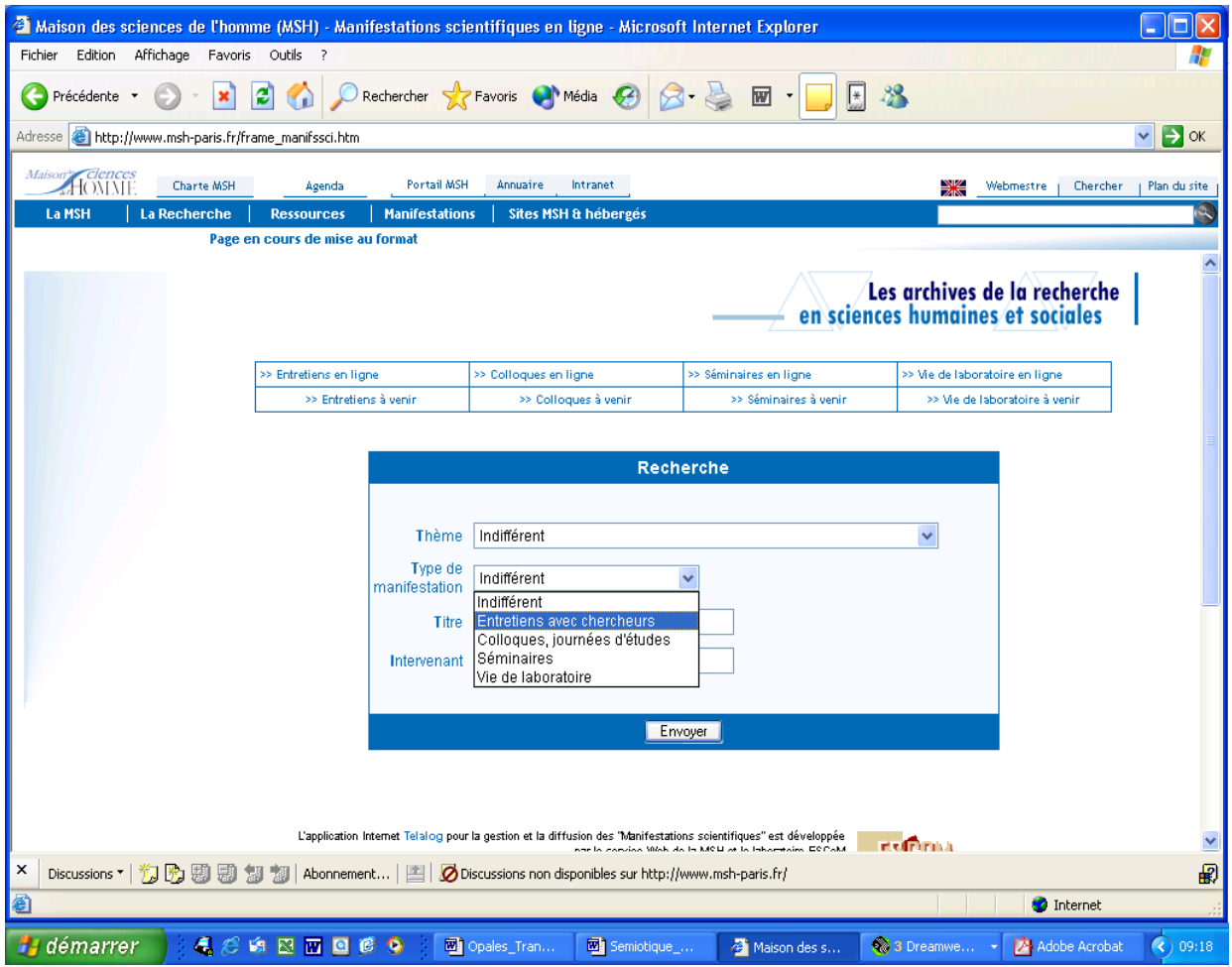

(figure 1 : the "simple" access page to the audiovisual archive in humanities of the Maison des Sciences de l'Homme. Web address: http://www.msh-paris.fr/frame_manifssci.htm). This access ceased to function in 2016. To access and explore the AAR scientific heritage, please use this link:
http://www.archivesaudiovisuelles.fr/fr/

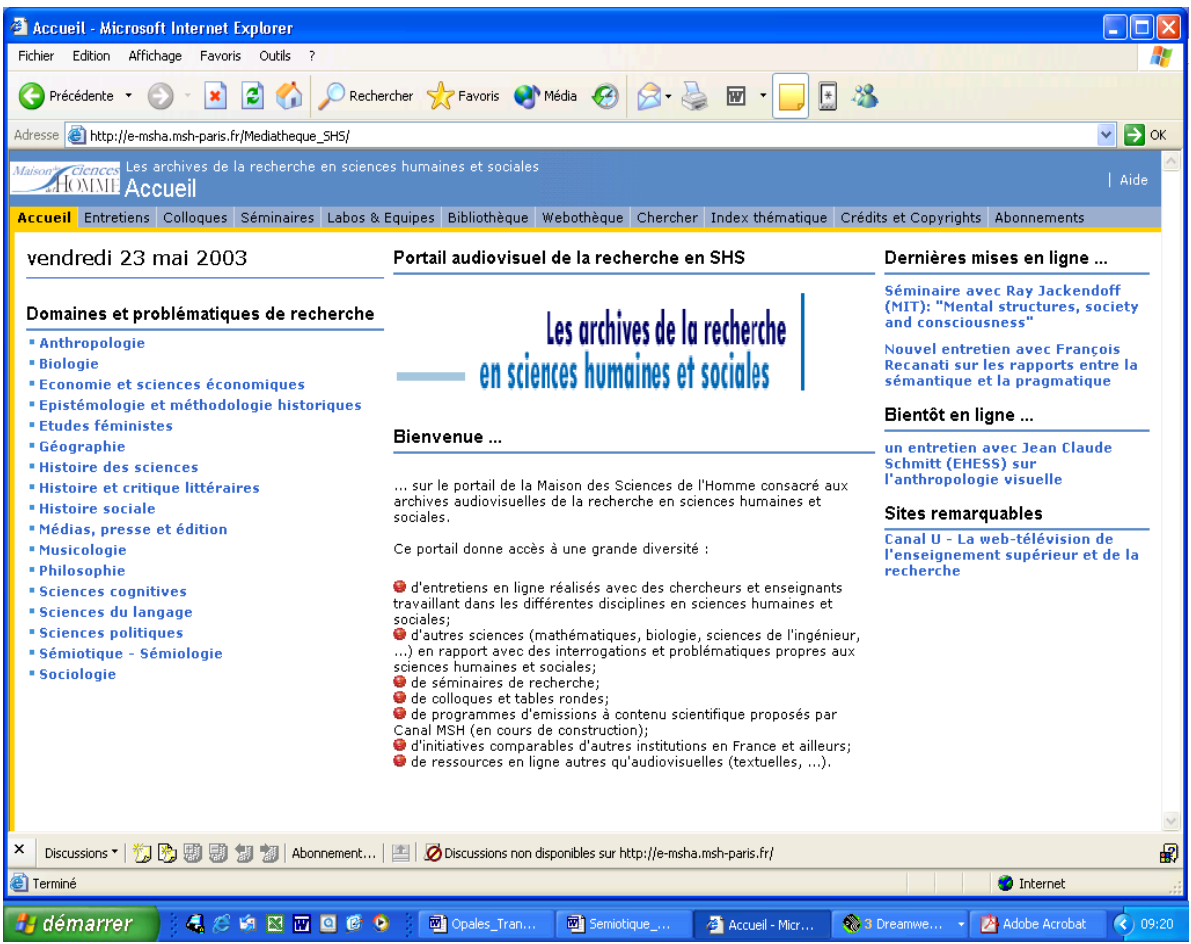

(figure 2: the web portal of the audiovisual archive in humanities of the Maison des Sciences de l'Homme. Web address: http://e-msha.msh-paris.fr/mediatheque_shs). This access ceased to function in 2016. To access and explore the AAR scientific heritage, please use this link: http://www.archivesaudiovisuelles.fr/fr/

**(figure 3a: a part of the "scientific events" library – the "Interviews")**

**(figure 3b: the description schema of the scientific event "Interview")**

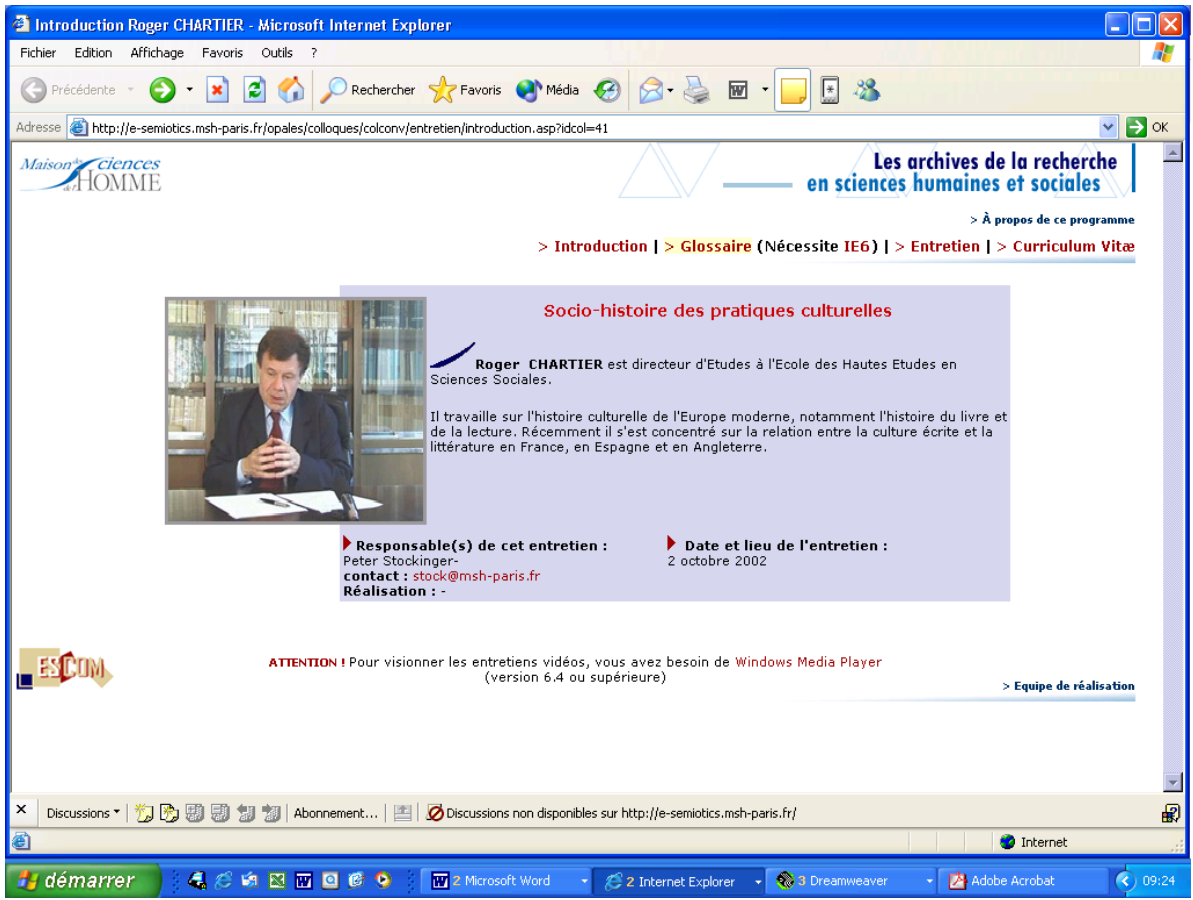

**(figure 4: "home page" of the dynamic website containing an interview with a researcher)**

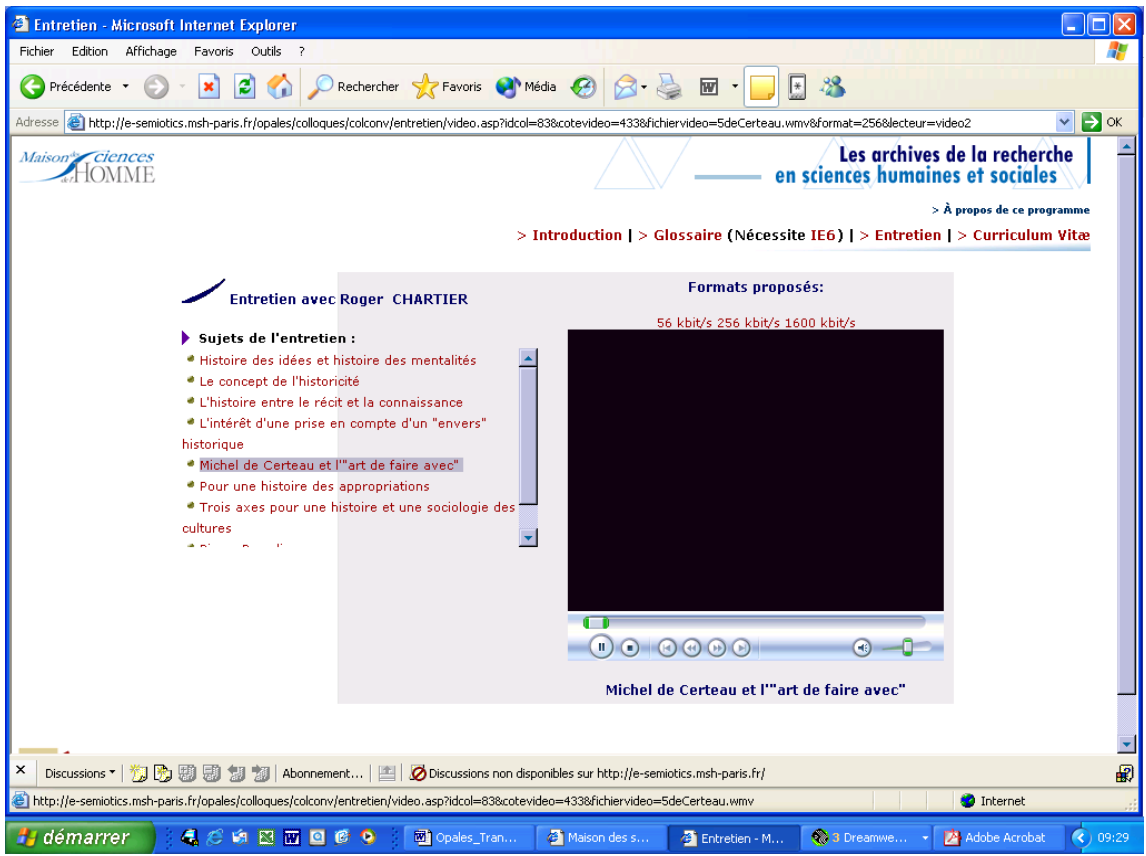

**(figure 5: the "video page" on the website containing an interview with a researcher)**

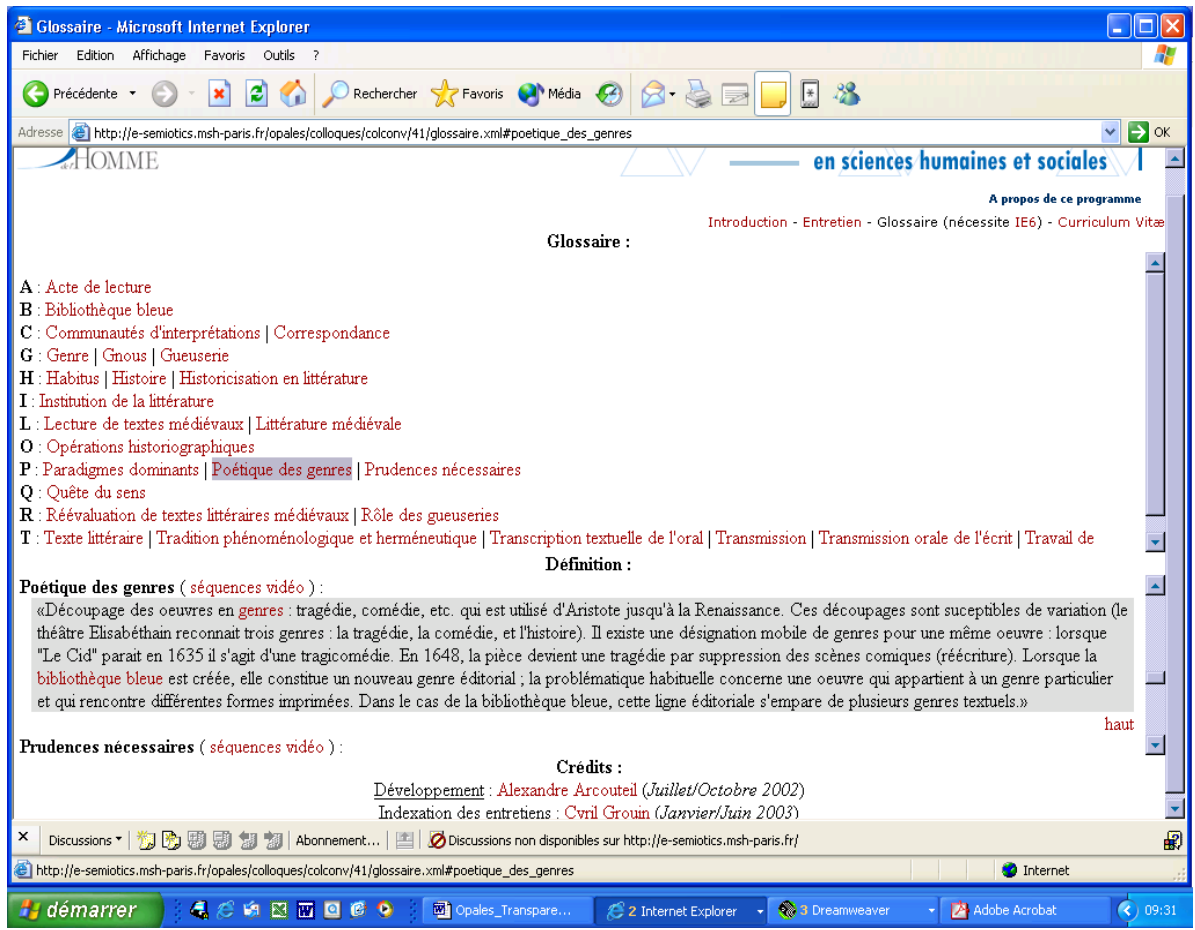

**(figure 6: the thematic index search and navigation interface of the website containing an interview)**

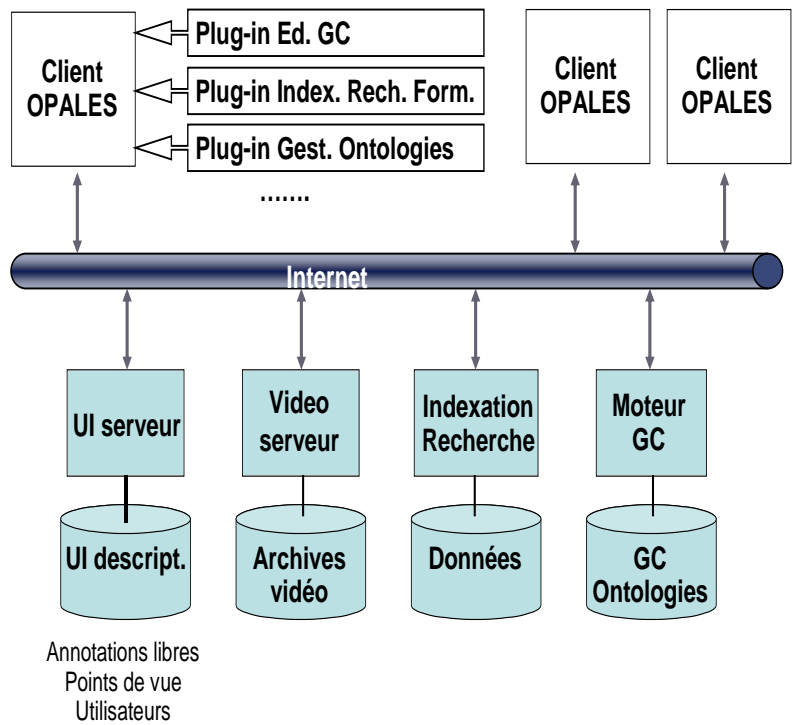

(figure 7 the general Opales architecture)

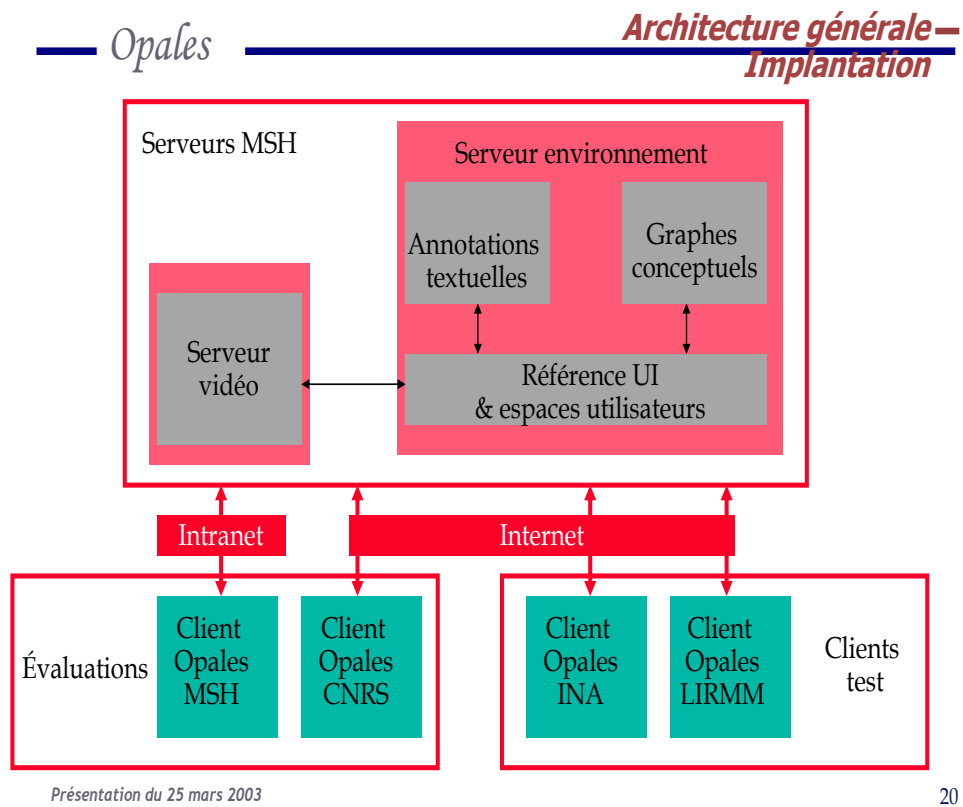

**(figure 8: the server – client architecture of Opales)**

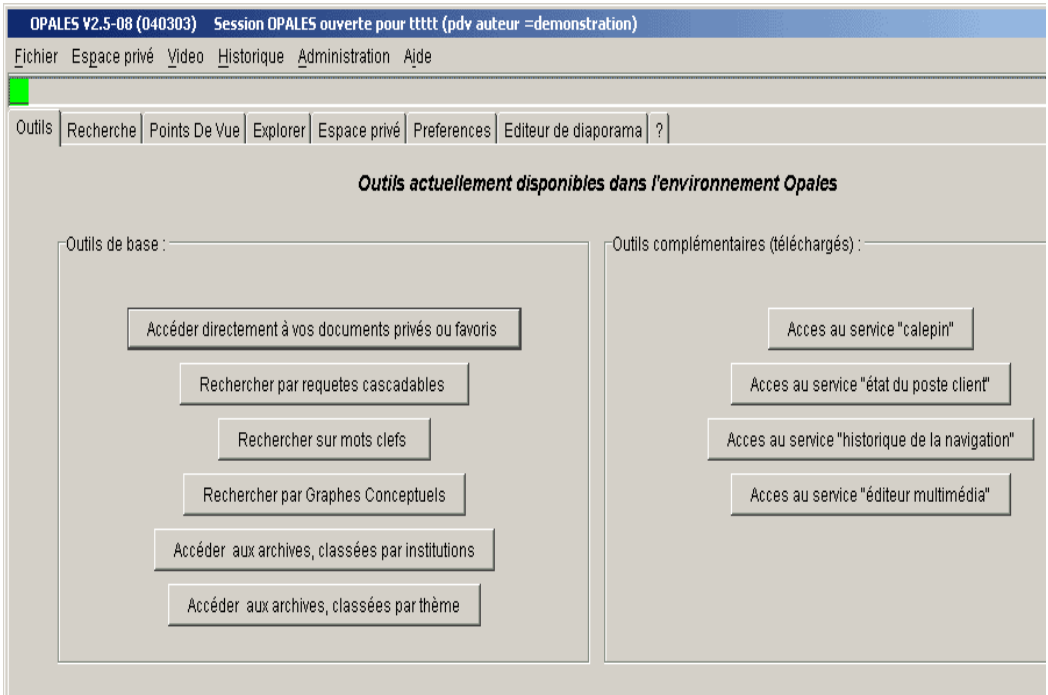

**(figure 9 : a sketch of the Opales search and working interface)**